\documentclass[12pt,a4paper,DIV12]{scrartcl}
\usepackage[T1]{fontenc}
\usepackage{lmodern}
\usepackage[british]{babel}
\usepackage{graphicx}
\usepackage{grffile}
\usepackage{hyperref}
\usepackage{color}
\usepackage{slashed}
\usepackage{textcomp}
\usepackage{subfigure}
\usepackage{amsmath}
\usepackage{amssymb}
\usepackage{amsfonts}
\usepackage{lscape}
\usepackage{psfrag}
\newcommand{\I}{\ensuremath{\mathrm{i}}}
\newcommand*{\email}[1]{\href{mailto:#1}{#1}} 
\newcommand{\arxiv}[2]{[arXiv:\,\href{http://arxiv.org/abs/#1}{\texttt{#1}} [\texttt{#2}]]}
\newcommand{\arxivold}[1]{[arXiv:\,\href{http://arxiv.org/abs/#1}{\texttt{#1}}\,]}
\newcommand{\aetap}{\text{a--}\eta'}
\newcommand{\api}{\text{a--}\pi}
\newcommand{\afn}{\text{a--}f_0}

\newcommand{\glg}{\tilde{g}g}
\newcommand{\su}[1]{\ensuremath{\text{SU}(#1)}}
\newcommand{\secref}[1]{Sect.~\ref{#1}}
\title{%
   {\vspace{-20mm}\normalsize
    \hfill\parbox[b][30mm][t]{35mm}{\textmd{MS-TP-15-29\\DESY 15-221}}}\\[-18mm]
The light bound states of supersymmetric SU(2) Yang-Mills theory\vspace*{2mm}}

\author{%
Georg Bergner\\
\textit{\large Albert Einstein Center for Fundamental Physics}\\
\textit{\large Institute for Theoretical Physics, University of Bern}\\
\textit{\large Sidlerstr.~5, CH-3012 Bern, Switzerland}\\
\textit{\large E-mail: \email{bergner@itp.unibe.ch}}\\[5mm]
Pietro Giudice, Gernot M\"unster\\
\textit{\large University of M\"unster, Institute for Theoretical Physics}\\
\textit{\large Wilhelm-Klemm-Str.~9, D-48149 M\"unster, Germany}\\
\textit{\large E-mail: \email{p.giudice@uni-muenster.de},
                       \email{munsteg@uni-muenster.de}}\\[5mm]
Istvan Montvay\\
\textit{\large Deutsches Elektronen-Synchrotron DESY}\\
\textit{\large Notkestr. 85, D-22603 Hamburg, Germany}\\
\textit{\large E-mail: \email{montvay@mail.desy.de}}\\[5mm]
Stefano Piemonte\\
\textit{\large University of Regensburg, Institute for Theoretical Physics}\\
\textit{\large Universit\"atsstr.~31, D-93040 Regensburg, Germany}\\
\textit{\large E-mail: \email{stefano.piemonte@ur.de}}
\vspace*{5mm}}

\date{December 22, 2015}
\begin{document}
\maketitle

\newpage

\begin{abstract}
\noindent
\textbf{\textsf{Abstract}}

Supersymmetry provides a well-established theoretical framework for
extensions of the standard model of particle physics and the general
understanding of quantum field theories. We summarise here our
investigations of $\mathcal{N}=1$ supersymmetric Yang-Mills theory with
SU(2) gauge symmetry using the non-perturbative first-principles method of
numerical lattice simulations. The strong interactions of gluons and their
superpartners, the gluinos, lead to confinement, and a spectrum of bound
states including glueballs, mesons, and gluino-glueballs emerges at low
energies. For unbroken supersymmetry these particles have to be arranged in
supermultiplets of equal masses. In lattice simulations supersymmetry can
only be recovered in the continuum limit since it is explicitly broken by
the discretisation. We present the first continuum extrapolation of the mass
spectrum of supersymmetric Yang-Mills theory. The results are consistent
with the formation of supermultiplets and the absence of non-perturbative
sources of supersymmetry breaking. Our investigations also indicate that
numerical lattice simulations can be applied to non-trivial supersymmetric
theories.
\end{abstract}

%%%%%%%%%%%%%%%%%%%%%%%%%%%%%%%%%%%%%%%%%%%%%%%%%%%%%%%%%%%%%%%%%%%%%%%%
\section{Introduction}

$\mathcal{N}=1$ supersymmetric Yang-Mills theory (SYM) is the supersymmetric
extension of the gluonic sector of the Standard Model. It contains
non-Abelian gauge fields of an \su{N} gauge group interacting with their
fermionic superpartners, the gluino fields. Different from the quarks of
QCD, the gluinos are Majorana fermions and they transform according to the
adjoint representation of the gauge group. The complexity of SYM is
comparable to QCD. Several basic properties, like asymptotic freedom, are
shared among these two theories~\cite{Amati:1988ft}. At low temperatures SYM
is assumed to confine the gluons and gluinos into colourless bound states,
similarly to the mesons and glueballs in QCD. Like in QCD, the investigation
of the bound state problem is a non-perturbative problem that can be
addressed by numerical lattice simulations.

Supersymmetry (SUSY) is the essential feature distinguishing SYM from QCD.
If SUSY is unbroken, the bound states are arranged in supermultiplets,
containing bosonic and fermionic particles with equal masses. The key aspect
of the investigations of SYM is to verify the formation of these multiplets
The obtained mass spectrum provides insights into the low energy effective
action of SYM. Effective actions have been constructed
in~\cite{Veneziano:1982ah} and extended
in~\cite{Farrar:1997fn,Farrar:1998rm}. Later the question about the
existence of SUSY breaking vacua has been raised in~\cite{Bergamin:2003ub},
which would imply a completely different mass spectrum. Another interesting
aspect is the existence of a stable light scalar in the theory, since in
addition to the fermion, the multiplet always contains a scalar and a
pseudoscalar particle. In SYM the scalar is therefore a natural component of
the effective theory, whereas the interpretation of its QCD counterpart is
more controversial~\cite{McNeile:2008sr}. A light scalar state is also
essential for technicolour and composite Higgs theories. The predicted
degeneracy of the multiplet masses provides a test of our methods for the
challenging measurements of this state.

An important characteristic of SUSY is the non-trivial interplay with the
space-time symmetries. For example, the anticommutator of the SUSY
generators $Q_\alpha$ is connected with the generators of translations
$P_\mu$:
\begin{equation}
\{Q_\alpha, Q_\beta\} = (C \gamma_\mu)_{\alpha\beta} P_\mu\,.
\end{equation}
The absence of the infinitesimal translations generated by $P_\mu$ is an
illustration of the unavoidable SUSY breaking on the lattice. In a more
detailed analysis one can prove that supersymmetry is generically broken on
the lattice~\cite{Bergner:2009vg}. In most cases there is no restoration of
SUSY in the continuum limit without a fine tuning of certain SUSY breaking
counterterms. In SYM it is the fine tuning of the bare gluino mass. Our
simulations are an important test for the general applicability of
non-perturbative lattice methods for SUSY theories. It is one of the few
non-trivial four-dimensional examples where the complete restoration of SUSY
in the continuum limit can be shown in terms of the Ward identities and the
degenerate mass spectrum.

The main focus of our investigations is the spectrum of bosonic and
fermionic bound states. In addition we have investigated the supersymmetric
Ward identities, static potential, thermal behaviour and other quantities.
Compared to QCD, the determination of the masses of meson-like bound states
in SYM is significantly more demanding, because all mesons are flavour
singlets and the calculation of their correlators requires the notoriously
difficult disconnected contributions. Like their QCD counterparts, the
glueball operators require a large statistics for a reasonable signal.

This article concludes our studies of the bound states spectrum of \su{2}
SYM,
see~\cite{Demmouche:2010sf,Bergner:2011wf,Bergner:2012rv,Bergner:2012eg,Bergner:2013nwa}
and references therein for previous work of our collaboration.
In~\cite{Demmouche:2010sf} a rather large gap between bosonic and fermionic
masses in the mass spectrum was obtained. In later
investigations~\cite{Bergner:2013nwa} we have found that this effect
significantly decreases at a smaller lattice spacing. In this article we
present the results at a third, even smaller, lattice spacing, which allow
an extrapolation to the continuum limit.

In \secref{sec:SYMC} we will give a short overview of the SYM in the
continuum. We describe the numerical setup for the lattice action and the
analysis in \secref{sec:NSim}. \secref{sec:b190} summarises our new data at
our smallest reliable lattice spacing. In combination with our previous
results the new data allow the continuum extrapolation of the spectrum as
presented in \secref{sec:extr}. Based on this extrapolation, in
\secref{sec:improv} we estimate the effects of improvements of the lattice
action that are interesting starting points for future investigations of the
theory.

%%%%%%%%%%%%%%%%%%%%%%%%%%%%%%%%%%%%%%%%%%%%%%%%%%%%%%%%%%%%%%%%%%%%%%%%
\section{$\mathcal{N}=1$ Supersymmetric Yang-Mills theory}
\label{sec:SYMC}

SYM has one conserved supercharge that transforms bosonic and fermionic
states into each other. In particular the gluino, a spin-$1/2$ Majorana
fermion in the adjoint representation of the gauge group \su{N}, is the
superpartner of the gluon.

SYM shows confinement at low energies and the gluons and gluinos form bound
states. In that respect it is similar to QCD, where the hadrons are formed
of the elementary quarks and gluons. On the other hand, SYM represents the
pure gluonic part of supersymmetric QCD and leads to the confinement of
external fundamental charges, corresponding to heavy fundamental quarks. The
linear rise of the static quark potential with a non-zero string tension
$\sigma$ is a measurable signal of this effect. Since the fermions are in
the adjoint representation, string breaking does not occur at any distance,
like in pure Yang-Mills theory. A deconfinement phase transition is expected
at a critical temperature $T_c$ that separates this low energy regime from
the gluino-gluon plasma at high temperature. First results have been
presented in~\cite{Bergner:2014saa}, a second order phase transition has
been observed at a temperature $T_c \simeq 200$ MeV in QCD units.

In the continuum the off-shell Lagrangian of SYM reads
\begin{equation}
\mathcal{L} = -\frac{1}{4} \textrm{Tr}( F_{\mu\nu}F^{\mu\nu} ) 
+ \frac{\I}{2} \bar{\lambda}(x)\gamma^{\mu} D_{\mu}\lambda(x) 
+ D^a(x) D_a(x),
\end{equation}
where $D_{\mu}$ denotes the covariant derivative in the adjoint
representation. The auxiliary field $D^a(x)$, needed to ensure supersymmetry
off-shell, has no kinetic term and can be integrated out from the partition
function. After adding a gluino mass term, that breaks SUSY softly, the
resulting on-shell Lagrangian is thus given by
\begin{equation}
\mathcal{L} = -\frac{1}{4} \textrm{Tr}( F_{\mu\nu}F^{\mu\nu} ) 
+ \frac{\I}{2} \bar{\lambda}(x) \gamma^{\mu} D_{\mu} \lambda(x)
-\frac{m_g}{2}\bar{\lambda}\lambda\,.
\end{equation}
In one-flavour QCD the action would look quite similar and the chiral
symmetry would correspond to $\mathrm{U}(1)_A \times \mathrm{U}(1)_V$ with a
$\mathrm{U}(1)_A$ broken by the anomaly and the unbroken $\mathrm{U}(1)_V$,
which corresponds to the conserved baryon number. In SYM the action is
invariant under a $\mathrm{U}(1)_R$ chiral symmetry that is broken down to
$Z_{N}$ by the anomaly. The formation of the gluino condensate leads to an
additional spontaneous breaking down to $Z_2$. The remaining symmetry
corresponds to the fermion number conservation modulo 2 for Majorana
fermions.

Based on low-energy effective actions, predictions have been made for two
low-lying chiral
supermultiplets~\cite{Veneziano:1982ah,Farrar:1997fn,Farrar:1998rm}. One of
them contains a scalar meson $\afn$, represented by the interpolating field
$\bar{\lambda}\lambda$, a pseudo-scalar meson $\aetap$, represented by
$\bar{\lambda} \gamma_5 \lambda$, and a gluino-glue state. The gluino-glue
is an exotic particle, which does not have a counterpart in QCD. It is a
spin 1/2 Majorana fermion, which can be created by the operator
\begin{equation}
\label{eq:gg}
\tilde{O}_{g\tilde{g}}=
\sum_{\mu\nu} \sigma_{\mu\nu} \textrm{Tr} \left[ F^{\mu\nu} \lambda\right],
\end{equation}
with $\sigma_{\mu\nu}=\frac{1}{2} \left[ \gamma_\mu,\gamma_\nu \right]$. The
other supermultiplet consists of a scalar $0^{++}$ glueball, a pseudoscalar
$0^{-+}$ glueball, and a gluino-glue particle.

%%%%%%%%%%%%%%%%%%%%%%%%%%%%%%%%%%%%%%%%%%%%%%%%%%%%%%%%%%%%%%%%%%%%%%%%
\section{Numerical lattice simulations}
\label{sec:NSim}

\subsection{Lattice formulations and simulation methods}

A lattice action for the Euclidean version of SYM has been proposed by Curci
and Veneziano~\cite{Curci:1986sm}. The action for the gauge fields is the
usual Wilson action, which in our calculations has been extended to the
tree-level Symanzik improved gauge action. The gluinos are described by
Wilson fermions in the adjoint representation. In our case stout
smearing~\cite{Morningstar:2003gk} is applied on the gauge links in the
Wilson-Dirac operator.

The Curci-Veneziano action explicitly breaks supersymmetry and the U(1)$_R$
symmetry. To recover the continuum symmetries the necessary fine-tuning of
supersymmetry and U(1)$_R$ symmetry can be achieved through the same
parameter, namely the bare gluino mass represented by the fermionic hopping
parameter $\kappa$~\cite{Curci:1986sm,Suzuki:2012pc}. To approach the limit
of a vanishing gluino mass, the bare parameter $\kappa$ has to be tuned to
the critical value $\kappa_c$ that corresponds to the point where all
explicit chiral symmetry breaking terms vanish. The value of $\kappa_c$ is
most easily obtained from the dependence of the adjoint pion ($\api$) mass
on $\kappa$. The correlator of this particle is the connected contribution
of the $\aetap$ correlator. Even though $\api$ is not a physical state of
SYM, it can be defined in a partially quenched setup where the chiral limit
is identified with the point where the adjoint pion mass
vanishes~\cite{Munster:2014cja}.

The updates of gauge configurations are performed with the two-step
polynomial hybrid Monte Carlo (PHMC)
algorithm~\cite{Montvay:2005tj,Demmouche:2010sf}. The polynomial
approximation is less precise for small eigenvalues of the Hermitian
Wilson-Dirac operator that can appear in the simulations for $\kappa$ close
to $\kappa_c$. When necessary, this error is corrected by a reweighting with
correction factors in the analysis. These are obtained from the exact
contribution of the lowest eigenvalues.

Our lattice formulation leads to a mild sign problem, arising as $O(a)$
lattice artefact. The Pfaffian obtained from the integration of the Majorana
fermions can sometimes have a negative sign~\cite{Bergner:2011zp},
especially close to the chiral limit. This sign is included in the
reweighting, if necessary. To reduce the statistical errors we have chosen
the parameters of our present simulations such that the reweighting with
correction factors and Pfaffian signs is not significant for the final
results. In general the relevance of the reweighting is reduced at the
smaller lattice spacings and for higher levels of stout smearing.

In principle it is also possible to formulate the model using
Ginsparg-Wilson type fermions. In such a formulation the parameter values,
at which the chiral and supersymmetric continuum limit is to be found, are
known and thus do not need fine tuning. Also, this formulation does not have
a sign problem of the Pfaffian. Nevertheless, there is supersymmetry
breaking at non-zero lattice spacings. Interesting results have been
obtained in some studies of the chiral condensate using this type of lattice
formulation~\cite{Giedt:2008xm,Endres:2009yp,Kim:2011fw}. The determination
of the mass spectrum requires, however, rather large lattices, which leads
to a high computational cost with these formulations. We have found in our
studies that the fine-tuning is feasible without problems, and our
investigations could be done without Ginsparg-Wilson type formulations.

%%%%%%%%%%%%%%%%%%%%%%%%%%%%%%%%%%%%%
\subsection{Simulation parameters}

\begin{figure}[ht]
\centering
\psfrag{AA}{$\langle \bar{\lambda}\lambda\rangle < 0$}
\psfrag{BB}{$\langle \bar{\lambda}\lambda\rangle > 0$}
\includegraphics[width=0.48\textwidth]{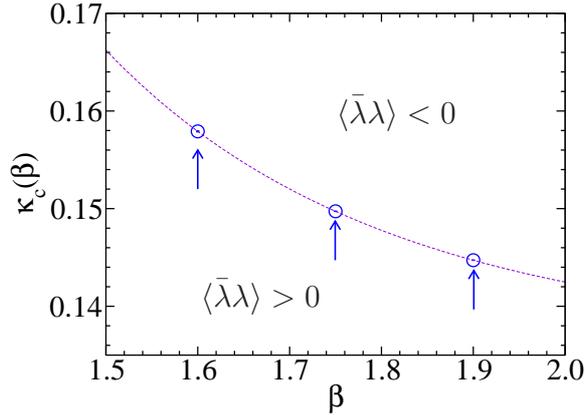}
\caption{Masses of particles are first extrapolated to the chiral limit,
i.e.\ to the critical value of $\kappa$ where the gluino mass vanishes, and
then to the continuum limit, $\beta\rightarrow \infty$. The critical line
$\kappa_c(\beta)$ separates the bare parameter space into two regions
characterised by either a positive or a negative expectation value of the
gluino condensate. Chiral extrapolations at $\beta=1.6$ and $\beta=1.75$
have been presented in Ref.~\cite{Demmouche:2010sf} and
Ref.~\cite{Bergner:2012rv,Bergner:2013nwa}, respectively.
\label{lcp_susy_summary}}
\end{figure}
We have chosen the lattice sizes such that finite volume effects can be
neglected in comparison with statistical errors, based on our investigation
in~\cite{Bergner:2012rv}. Since lattice artifacts are the most relevant
systematic errors, the most reliable results would be obtained at the
smallest lattice spacings. The lattice spacing can, however, not be easily
reduced to arbitrary small values since topology freezing introduces very
large autocorrelation times and uncertainties in the observables at a small
lattice spacing~\cite{Luscher:2011kk}. This problem can be reduced by
choosing longer HMC trajectories in the gauge field update, for instance
trajectories of length 2, which we had in most of our simulations at larger
$\beta$ values. In spite of that we have found a significant influence of
topology freezing at $\beta=2.1$, where the lattice spacing in QCD units
(where the Sommer parameter $r_0$ is set to 0.5 fm) would be around $0.02$
fm. For instance, at $\beta=2.1,\ \kappa=0.1397$ on a $48^3 \cdot 96$
lattice we observe for the topological charge $Q$ an integrated
autocorrelation time $\tau_Q \simeq 145$, whereas at $\beta=1.9,\
\kappa=0.14415$ on a $32^3 \cdot 64$ lattice we have $\tau_Q \simeq 24$. In
addition to the longer autocorrelation, at $\beta=2.1$ it is very difficult
to achieve a distribution of $Q$ symmetric about 0. Topology freezing
represents an upper limit for the parameter $\beta$ in the simulations
unless open boundary conditions are introduced or the statistic is increased
by about an order of magnitude.

In this work we present our new results of simulations at $\beta = 1.9$
which are summarised in Tab.~\ref{tab:chirallimit190} and discussed in more
detail in \secref{sec:b190}. Including our previous simulations, three
different values of the inverse gauge coupling $\beta$, corresponding to
three lattice spacings $a$ have been considered. For each $\beta$, several
values of the fermionic hopping parameter $\kappa$ have been chosen in order
to extrapolate to $\kappa_c$, see Fig.~\ref{lcp_susy_summary}. The
parameters of the simulations at the two larger lattice spacings at $\beta =
1.6$ and $1.75$ have been presented in previous
publications~\cite{Demmouche:2010sf,Bergner:2012rv,Bergner:2013nwa}.

%%%%%%%%%%%%%%%%%%%%%%%%%%%%%%%%%%%%%
\subsection{Particle operators on the lattice}

The mesonic operators are similar to the flavour singlet meson operators in
QCD. The scalar meson corresponds to the adjoint version of $f_0$ ($\afn$),
and the pseudoscalar to the adjoint $\eta'$ meson ($\aetap$). The
correlators for these particles are a sum of disconnected and connected
fermion contributions. In QCD the connected part corresponds to the pion,
while in SYM an adjoint pion ($\api$) is not a physical state of the theory,
but can be defined in a partially quenched setup~\cite{Munster:2014cja}. It
provides the best signal to noise ratio of all the considered states and the
most reliable basis for determining the critical value $\kappa_c$ of the
hopping parameter where the gluino mass vanishes.

The determination of the disconnected contributions is rather challenging.
They are determined by a stochastic estimation combined with the exact
contribution of the lowest eigenvalues of the Hermitian Dirac-Wilson
operator and truncated solver techniques to reduce the fluctuations of the
signal. Similar to QCD, the disconnected contributions yield the most
relevant uncertainty in the mesonic operators.

The gluino-glue is measured with a lattice version of the operator of
Eq.~\eqref{eq:gg}, where the $F_{\mu\nu}$ part is replaced by the clover
plaquette. APE and Jacobi smearing is applied to get a better signal for the
ground state~\cite{Bergner:2012rv}.

The glueball masses are determined on the lattice by operators based on the
product of link variables. In our study the interpolating operator for the
scalar glueball $0^{++}$ is given by the fundamental plaquette built from
four links, while for the glueball $0^{-+}$ it is given by the product of
eight links with suitable shape~\cite{Kamel:2009}.

In order to reduce the contamination from excited states and therefore
determine the effective mass already at small time-slice separation we used
the variational method based on APE smeared
operators~\cite{Albanese:1987ds}. In total between $L=16$ and $L=18$
different operators were used in the variational method, each separated by
$N_{APE}$ steps. The smearing parameter was usually fixed to
$\epsilon_{APE}=0.5$, while $N_{APE}=4$ for the volume $24^3$ and
$N_{APE}=5$ for the volume $32^3$. As in QCD, the glueballs are
characterised by a small signal-to-noise ratio compared to the other
observables.

%%%%%%%%%%%%%%%%%%%%%%%%%%%%%%%%%%%%%%%%%%%%%%%%%%%%%%%%%%%%%%%%%%%%%%%%
\subsection{Supersymmetric Ward identities}

An important issue in our approach is the determination of the point in
parameter space where the theory is characterised by a massless gluino. In
this point not only the explicit chiral symmetry breaking by the gluino mass
disappears, but also, in the continuum limit, SUSY is restored.

A renormalised gluino mass can be defined on the lattice by means of the
supersymmetric Ward identities (SWI)~\cite{Farchioni:2001wx}. They give the
(subtracted) gluino mass up to a renormalisation factor ($am_S Z^{-1}_S$).

On the other hand, the point of vanishing gluino mass can be estimated in an
indirect way from the vanishing of the adjoint pion mass. The adjoint-pion
mass squared $\api$ is expected to vanish linearly with the (renormalised)
gluino mass $m^2_{\api} \propto m_g$. This relation was obtained in the OZI
approximation in~\cite{Veneziano:1982ah}, and derived
in~\cite{Munster:2014cja} in a partially quenched setup.

The parameter which we tune to get a zero gluino mass is $\kappa$. In
general, the $\api$ mass yields a more precise determination of $\kappa_c$.
In principle, different definition of the renormalised gluino mass will
agree up to $O(a)$ lattice artefacts. In previous studies we have checked
that both signals lead within our statistical errors to a consistent value
of $\kappa_c$~\cite{Demmouche:2010sf}.

\begin{figure}[ht]
\centering
\subfigure[$\beta=1.75$, volume $32^3\times 64$.]%
{\includegraphics[width=0.48\textwidth]{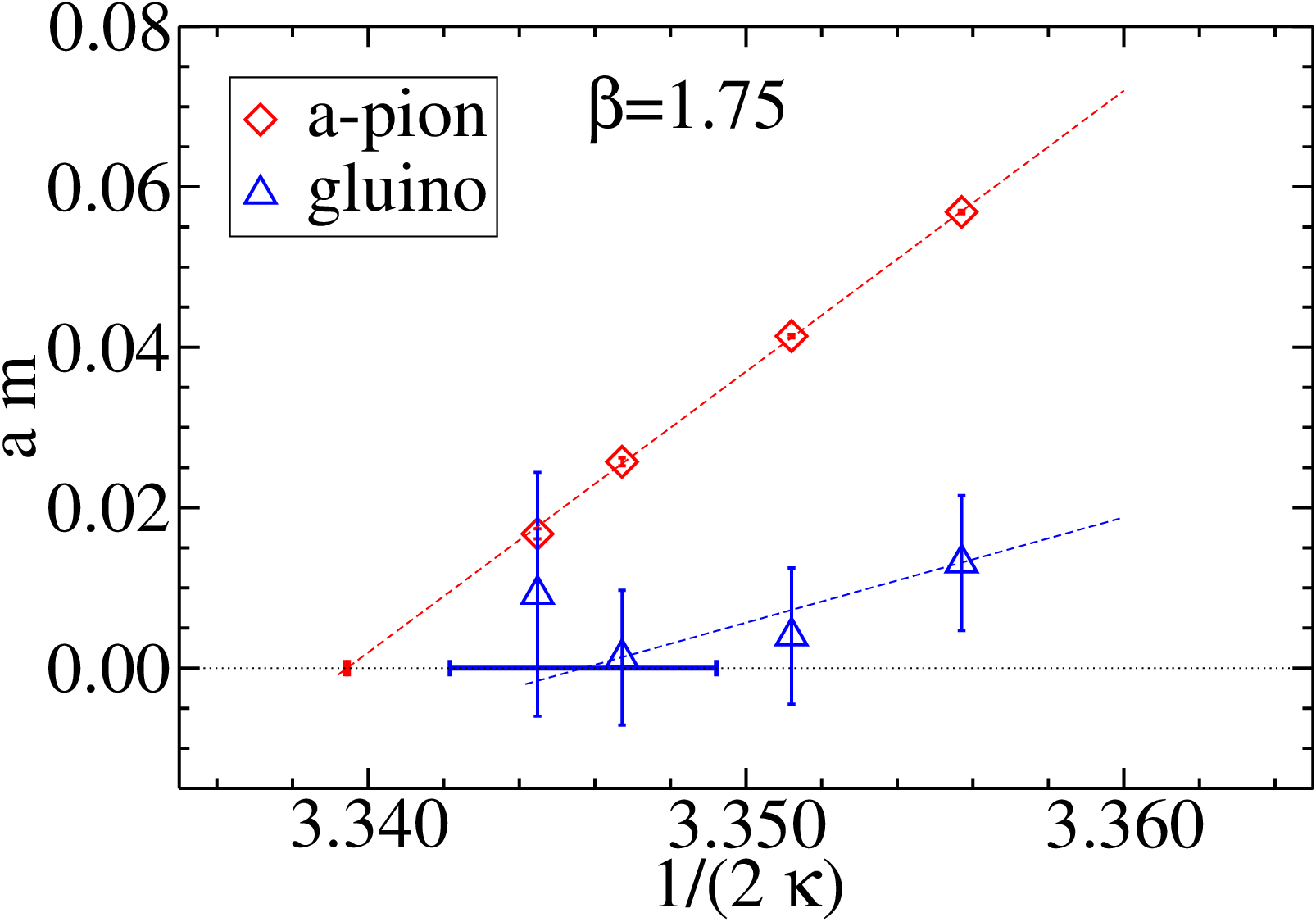}}
\subfigure[$\beta=1.90$, volume $32^3\times 64$.]%
{\includegraphics[width=0.50\textwidth]{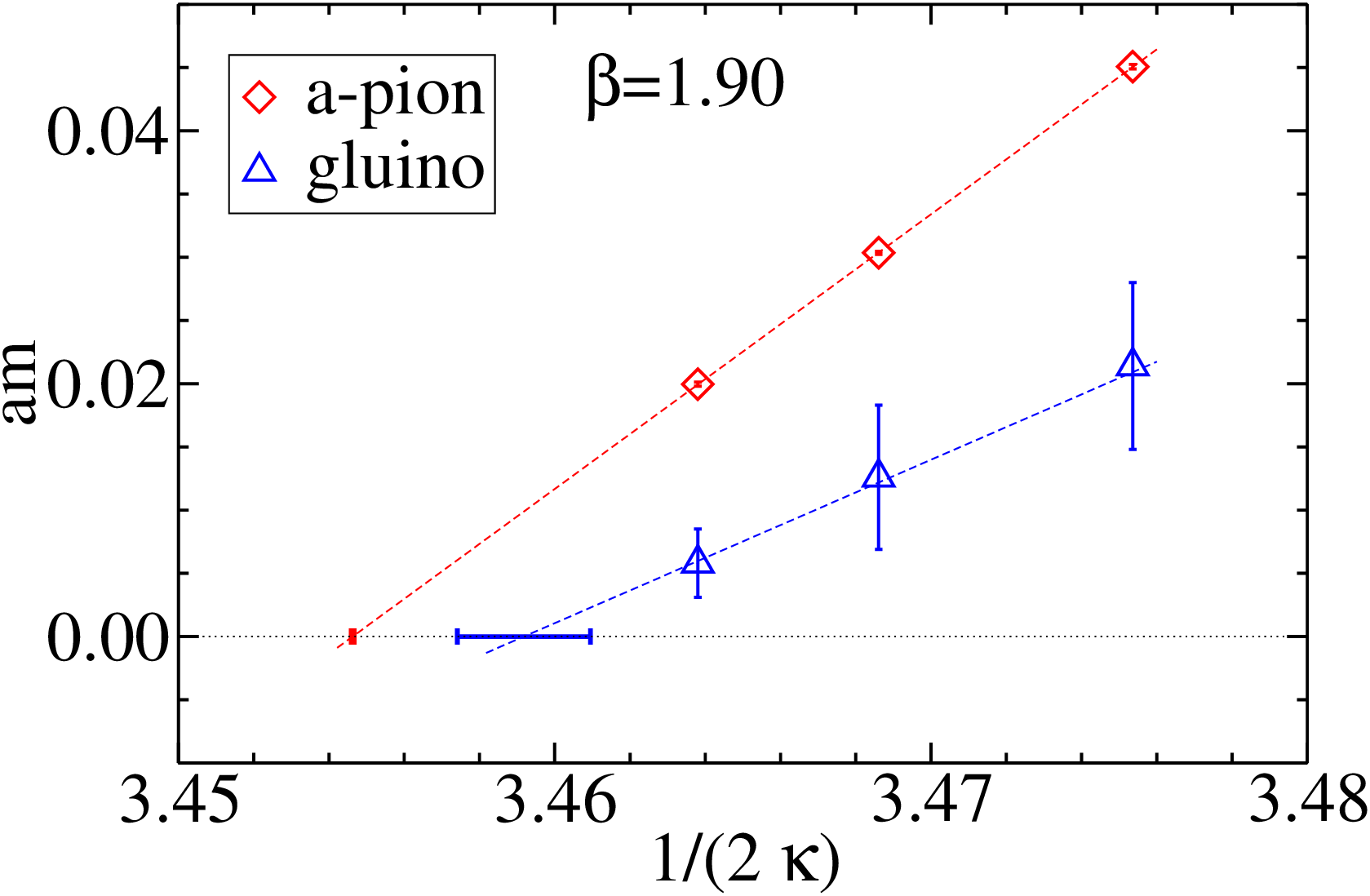}}
\caption{Comparison between the values of $\kappa_c$ defined as
the value of $\kappa$ where the square of the $\api$ mass or the
renormalised gluino mass $am_S Z^{-1}_S$ vanishes. The horizontal
thick lines (red and blue) show the uncertainty in the determination of the
intercept with zero.
\label{plot:wardidentities}}
\end{figure}

We have measured the SWI in our new set of configurations, closest to the
continuum limit, at $\beta=1.75$ and at $\beta=1.9$, to check again the
agreement in the determination of $\kappa_c$. In
Fig.~\ref{plot:wardidentities} we plot the value of $am_S Z^{-1}_S$
(labelled as gluino) and the square of the adjoint pion mass (labelled as
a-pion).

We can see that the two values are compatible within two standard deviations
both at $\beta=1.75$ and $\beta=1.9$. Taking into account that the
determination of $am_S Z^{-1}_S$ is a complex procedure, it is difficult to
estimate it's systematic errors, and we consider both determinations to be
acceptable and the two methods to be compatible.

Because the value of $\kappa_c$ determined via $\api$ is by far more
precise, the tuning of $\kappa$ is done using this quantity.

%%%%%%%%%%%%%%%%%%%%%%%%%%%%%%%%%%%%%%%%%%%%%%%%%%%%%%%%%%%%%%%%%%%%%%%%
\section{New results at $\beta=1.9$}
\label{sec:b190}

To estimate the masses of the bound states in the continuum at zero gluino
mass a two-fold extrapolation has to be made. In the first step, at each
fixed value of $\beta$ the masses are extrapolated in $\kappa$ to the limit
of a vanishing gluino mass at $\kappa_c$. As explained above, this is done
by considering the masses as a function of the squared mass $m_{\api}$ of
the adjoint point and extrapolating to the ``chiral limit'' $m_{\api} = 0$.
We consider a mass independent scale setting scheme, so that the lattice
spacing is considered to be constant at fixed $\beta$. Therefore the
extrapolations to the chiral limit can be done directly for the masses in
lattice units ($am$).

\begin{figure}[ht]
\centering
\subfigure[Gluino-glue and $\aetap$ mass.]%
{\includegraphics[width=0.48\textwidth]{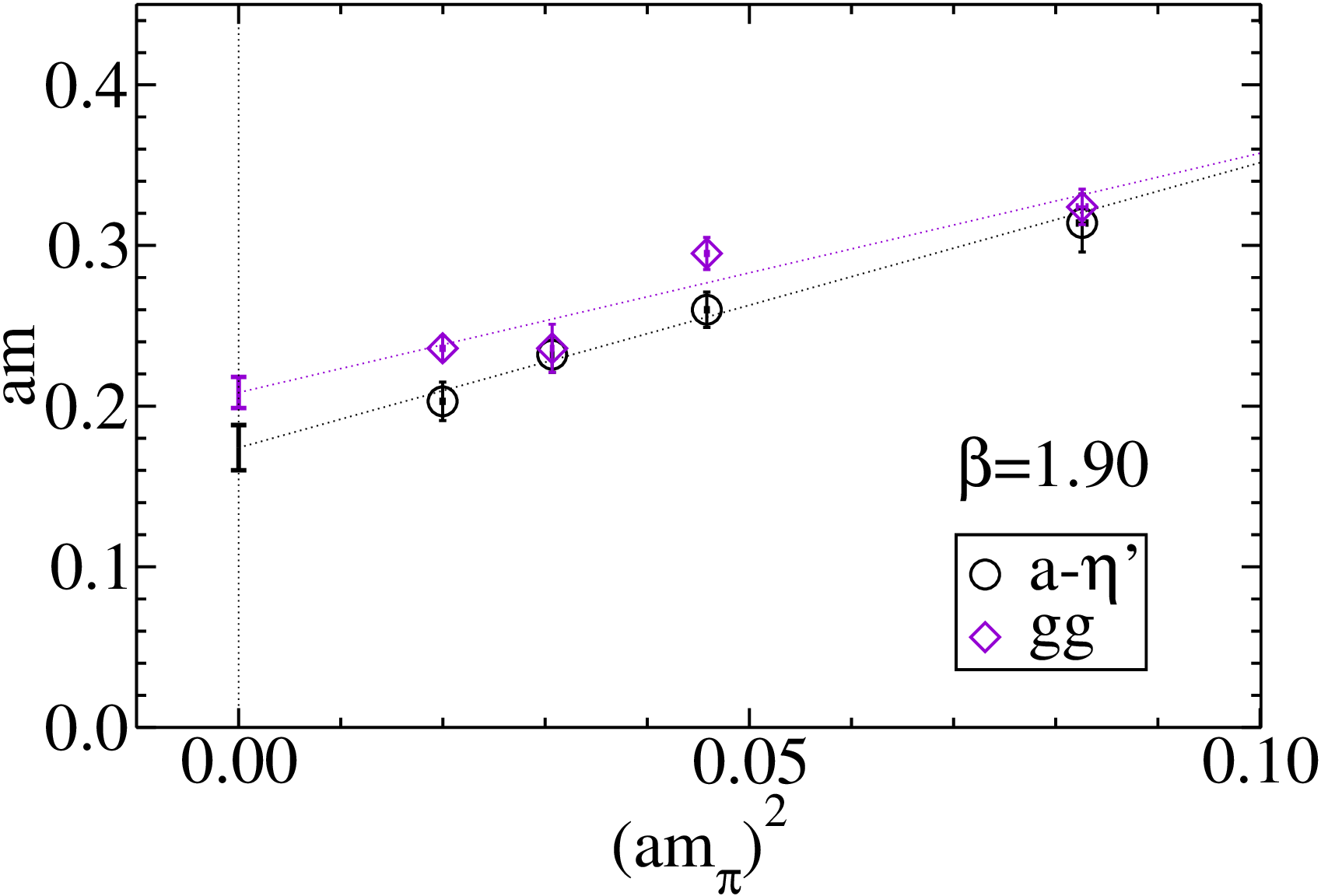}}
\subfigure[$0^{++}$ glueball and $\afn$ mass.]%
{\includegraphics[width=0.48\textwidth]{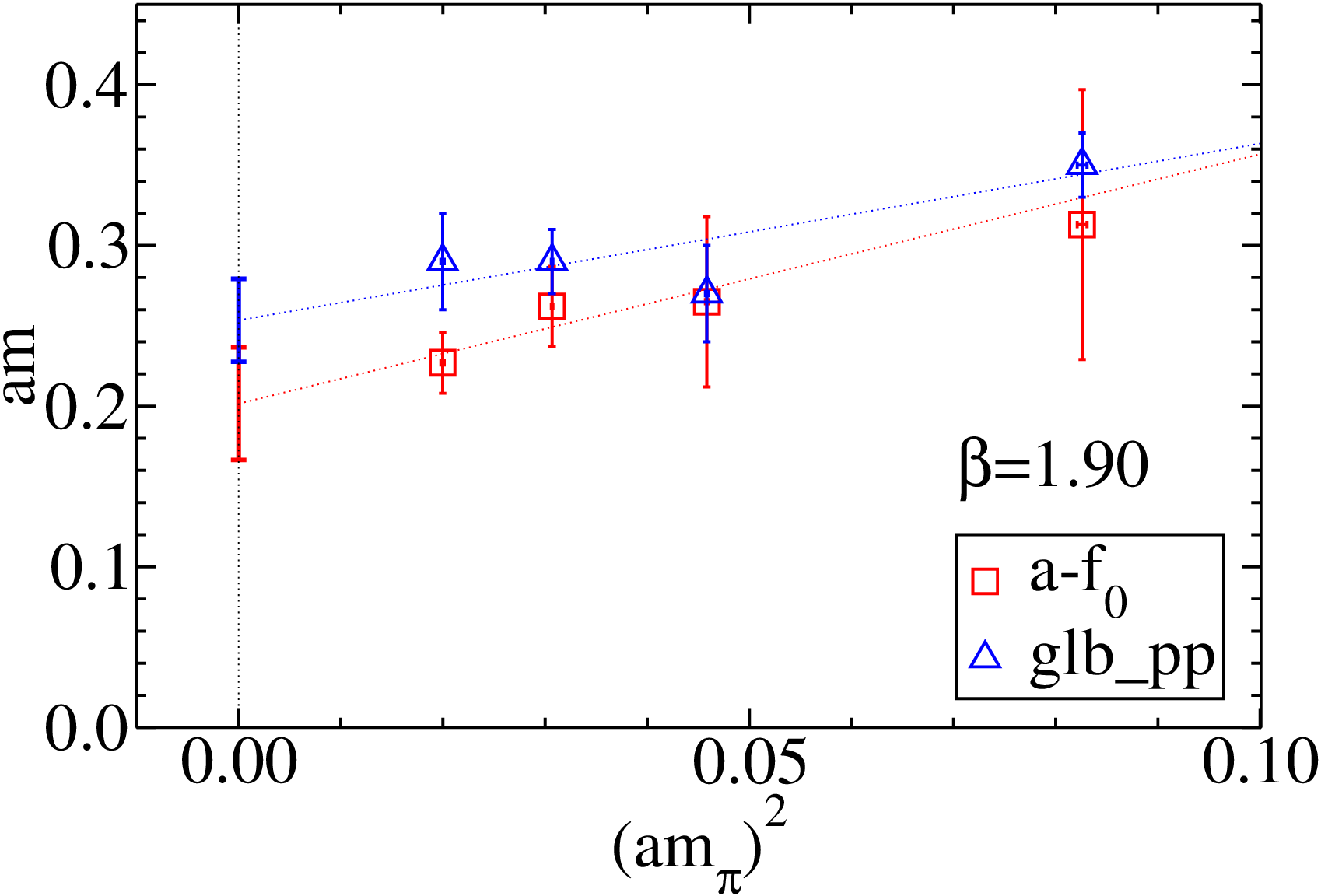}}
\caption{Extrapolation to the chiral limit at $\beta=1.90$.
\label{beta190chirallimit}}
\end{figure}

The masses in lattice units, obtained from our simulations at $\beta=1.9$,
are presented in Tab.~\ref{tab:chirallimit190}.
Fig.~\ref{beta190chirallimit} shows the extrapolation of the masses to the
chiral limit. As for the $\beta=1.75$ case, the spectrum is almost
degenerate in the chiral limit, which is quite different from our previous
results at $\beta=1.6$. This qualitative observation can now be made more
rigorous by a complete extrapolation to the continuum limit.

%%%%%%%%%%%%%%%%%%%%%%%%%%%%%%%%%%%%%%%%%%%%%%%%%%%%%%%%%%%%%%%%%%%%%%%%
\section{Extrapolations to the continuum limit}
\label{sec:extr}

\subsection{Low-lying masses}

The first extensive studies of our collaboration at the lattice spacing $a$
corresponding to $\beta=1.6$ have found a mass of the gluino-glue particle
$m_{\glg}$ that was significantly heavier than the mass $m_{\aetap}$ of the
$\aetap$-meson~\cite{Demmouche:2010sf}. In a second step $a$ has been
reduced, corresponding to a larger value of $\beta=1.75$, and a significant
reduction of the gap between $m_{\glg}$ and $m_{\aetap}$ has been
observed~\cite{Bergner:2012rv,Bergner:2013nwa}. The new results at the third
smaller lattice spacing at $\beta=1.9$ now allow to present the first
extrapolation to the continuum limit of the lowest bound state masses. The
large mass splitting between the bosonic bound states and their fermionic
counterpart that was visible at $\beta=1.6$ has been significantly reduced
at the smaller lattice spacings.%, see Fig.~\ref{chart}.

Results at finite lattice $a$ can be extrapolated to the continuum limit $a
\rightarrow 0$ once an observable is chosen to set the scale, i.~e.\ to
define the physical length associated with $a$. The scale setting is a large
source of systematic errors, therefore several different observables have
been computed for the extrapolation to the continuum limit. The
determination of the scale for our model has been presented in
Ref.~\cite{Bergner:2014ska}. To compute the continuum limit extrapolations,
we have chosen the Wilson flow quantity $w_0$ defined at the reference time
$\tau=0.3$. The lattice spacing is implicitly defined by the obtained
numerical value of the dimensionless quantity $a/w_0$. The Sommer parameter
$r_0$ could be a valid alternative, it has, however, more systematic
uncertainties than $w_0$, since it requires complex fitting procedures of
the noisy expectation values of Wilson loops. The corresponding results are
listed in Tab.~\ref{tab:chirallimit} for the three values of $\beta$. Our
current method for setting the scale is different from our previous studies,
where $r_0$ has been chosen to set the scale.

All chiral extrapolations have now been redone based on bare quantities in
lattice units as in \secref{sec:b190}. The error of the scale setting does
hence not propagate into the chiral extrapolations, which leads to smaller
errors. At $\beta=1.75$ we have now split the chiral fit into the ensembles
with one and three levels of stout smearing, even though the results are
compatible within errors. In that way the complete extrapolation is now
based only on data obtained with one level of stout smearing.

We have also redetermined the glueball masses $\beta=1.6$ with a more
accurate variational analysis. This leads to a reduced difference between
the gluino-glue and the glueball compared to our previous
publication~\cite{Demmouche:2010sf}.

\begin{figure}[ht]
\centering
\subfigure[Gluino-glue and $\aetap$ mass.]%
{\includegraphics[width=0.48\textwidth]{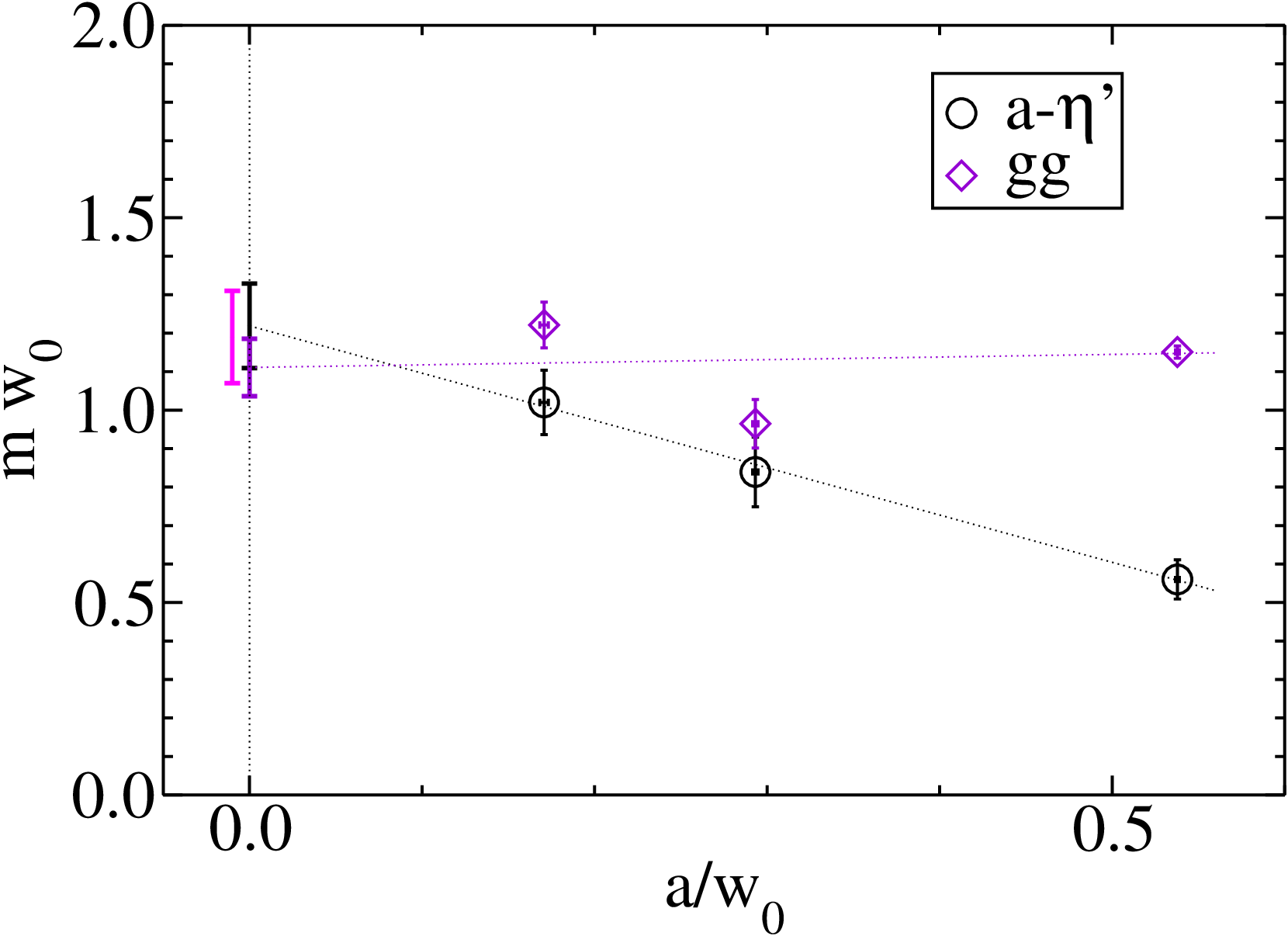}
\label{f0etacontinuumcomp}}%
\hspace{2mm}
\subfigure[$0^{++}$ glueball and $\afn$ mass.]%
{\includegraphics[width=0.48\textwidth]{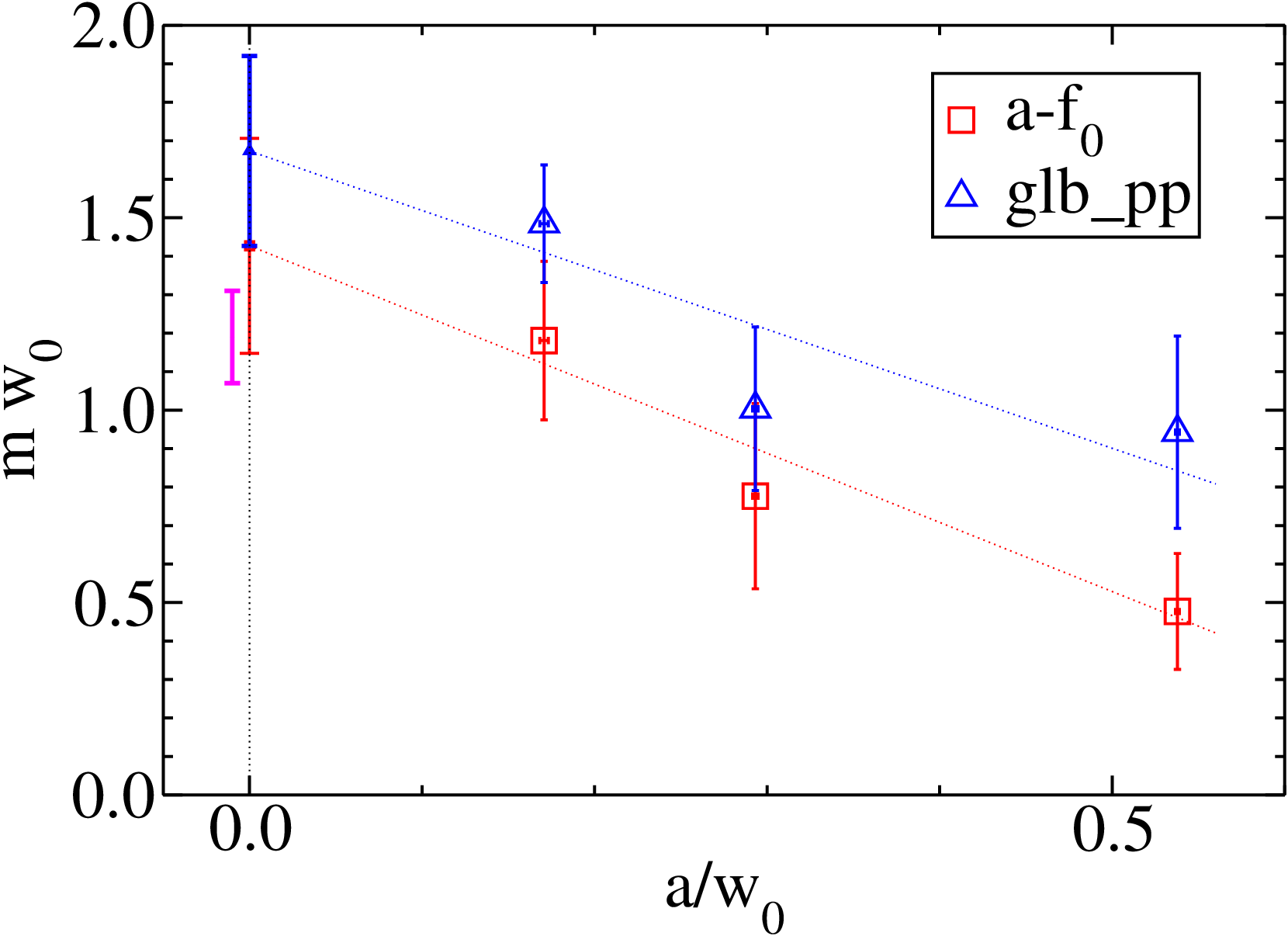}
\label{zppcontinuumcomp}}
\caption{a) Extrapolations of the $\aetap$ and gluino-glue masses to the
continuum limit. The large gap visible at the largest lattice spacing is
drastically reduced at larger $\beta$.
b) Extrapolations of the $\afn$ and $0^{++}$ glueball masses to the
continuum limit. The extrapolation of the glueball mass is unstable and
dominated by the result at small lattice spacing, therefore the final result
is compatible with the gluino-glue mass only within two standard deviations.
The thick vertical magenta line represents the weighted mean value of the four
masses.}
\end{figure}

The extrapolations of the masses to the continuum, using $w_0$ for the scale
setting, are displayed in Figs.~\ref{f0etacontinuumcomp} and
\ref{zppcontinuumcomp}.

Using $w_0$ to set scale, the final extrapolation to the continuum limit can
be read in Table~\ref{tab:continuumlimit}, labelled as $s=w_0$:
\begin{eqnarray}
w_0\, m_{\aetap} & = & 1.22(11) \\
w_0\, m_{\afn} & = & 1.43(28)\\
w_0\, m_{\glg} & = & 1.111(74) \\
w_0\, m_{\text{glueball}\ 0^{++}} & = & 1.67(25) \,.
\end{eqnarray}

The masses of $\aetap$, $\afn$ and $\glg$ are compatible with the weighted
mean value $w_0\, \bar{m}= 1.19(12)$ of the four masses within one standard
deviation, while the mass of the $0^{++}$ glueball is compatible with it by
only two standard deviations. However, the glueball has the worst
signal-to-noise ratio and the extrapolation of its mass to the continuum
limit is less reliable due to the large errors at fine lattice spacing.

%%%%%%%%%%%%%%%%%%%%%%%%%%%%%%%%%%%%%
\subsection{Glueballs}

\begin{figure}[htb]
\centering
\includegraphics[width=0.48\textwidth]{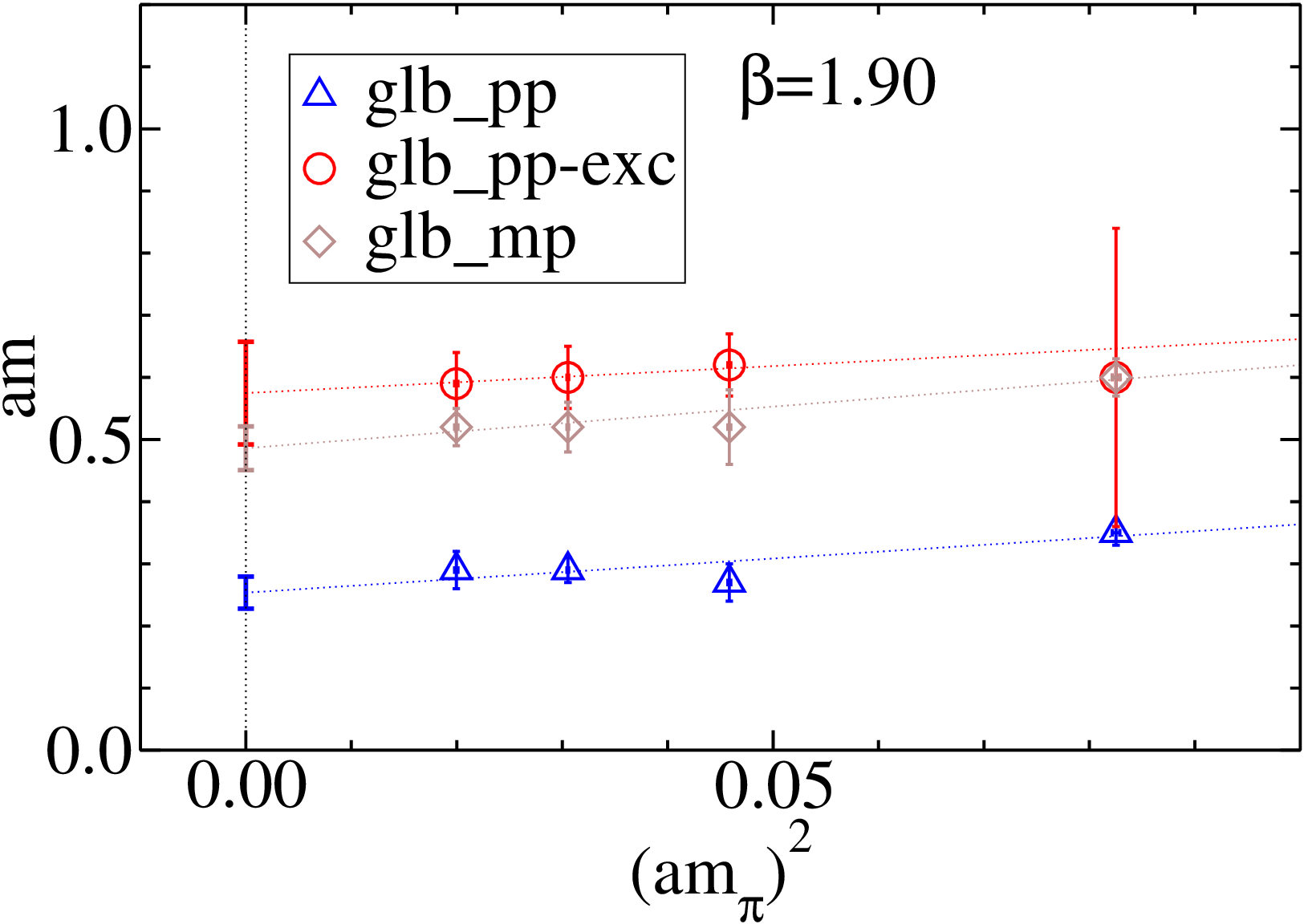}
\caption{Glueball spectrum at $\beta=1.9$. The mass of the first excited
state of the $0^{++*}$ glueball, which is twice the mass of the
$0^{++}$ glueball ground state, appears to be compatible with the
fundamental mass of the $0^{-+}$.
\label{beta190glueballspectrum}}
\end{figure}

The picture of a lower supermultiplet consisting of bound states of gluons
and a higher one of mesons was proposed in~\cite{Farrar:1998rm}. On the
other hand, in~\cite{Feo:2004mr} the authors, using different arguments and
experience from QCD, deduce the opposite ordering of multiplets. The
$0^{++}$ glueball and the gluinoball $\afn$ on the one hand, and the
$0^{-+}$ glueball and the gluinoball $\aetap$ on the other hand have the
same quantum numbers, and they might be characterised by a strong mixing. In
that case it would be difficult to say which of the two supermultiplets is
more glueball-like or gluino-like, since we will get the same mass from both
of them, namely the one of the lightest state. This is the case shown in
Fig.~\ref{zppcontinuumcomp}, where $0^{++}$ and $\afn$ have compatible
masses.

On the other hand, if the mixing of the states is weak, it is possible that
the two operators project onto two different states, and two different
masses are obtained in this case. Our results seem to suggest that the
$0^{-+}$ glueball and the $\aetap$ have a weak mixing, and the operators we
are using for the $0^{-+}$ glueball have a strong overlap with the first
excited state, but a very weak one with lowest state.

{}From Fig.~\ref{beta190glueballspectrum} we can see moreover that the first
excited state of the $0^{++}$ glueball is compatible with the lightest state
of the $0^{-+}$ glueball. In combination with an additional excited
gluino-glue, this would be a multiplet of excited states. This line of
reasoning seems to indicate that the glueball states have an energy higher
than the gluinoball as argued in~\cite{Feo:2004mr}.

%%%%%%%%%%%%%%%%%%%%%%%%%%%%%%%%%%%%%
\subsection{Comparison between $r_0$ and $w_0$}

The lattice spacing has been defined in terms of $w_0$ in the extrapolation
to the continuum limit. It could be interesting to express the final results
for the masses also in terms of the Sommer parameter $r_0$. There are two
possible ways to get these results: on the one hand the dimensionless ratio
$r_0/w_0$ can be extrapolated to the continuum limit and used to convert the
above continuum results to the scale $r_0$; on the other hand the
extrapolations to the continuum limit can alternatively be done using the
value of $a/r_0$ as implicit definition of the lattice spacing. The two
different procedures can be used to test the systematic error of the
extrapolation to the continuum limit. In fact, assuming that the asymptotic
scaling region has already been reached, the two procedures must give
compatible results. If not, this would be a signal of the fact that the
extrapolation to the continuum limit is not yet stable.

Using $r_0$ as scale parameter to extrapolate to the continuum limit we get
the results labelled as $s=r_0$ in Table~\ref{tab:continuumlimit}. The
dimensionless ratio $r_0/w_0$ extrapolated to the continuum limit is
\begin{equation}
r_0/w_0 =  2.21(12) \,.
\end{equation}
Using this value to convert the masses, which have been extrapolated to the
continuum limit using $r_0$, in terms of $w_0$ we get the values labelled
with $w_0^*$ in Table~\ref{tab:continuumlimit}. We note that for $\beta=1.6$
we have recalculated the Sommer parameter $r_0$ in order to apply a uniform
methodology for all values of $\beta$. The results for $r_0$ are slightly
different from the old ones in~\cite{Demmouche:2010sf}, since by a refined
choice of fit ranges we could obtain more reliable results.

For all particles, the masses determined with these two procedures are
compatible within 1.4 standard deviations. However, the use of $r_0$ results
in a slightly larger mass gap between the gluino-glue and the $\aetap$
masses. Other possible choices to set the scale, like $t_0$ or $w_0$ at a
different reference value, give an almost perfect compatibility, related to
the fact that they do not refer to completely independent observable.

%%%%%%%%%%%%%%%%%%%%%%%%%%%%%%%%%%%%%%%%%%%%%%%%%%%%%%%%%%%%%%%%%%%%%%%%
\section{Improved lattice formulations}
\label{sec:improv}

In the continuum extrapolation we have found a particle spectrum compatible
with the formation of a supermultiplet and unbroken supersymmetry. This is
also compatible with theoretical considerations that have found a non-zero
Witten index of the theory. We can now invert the argument and take the
separation of the multiplets as a signal for the lattice artefacts of the
chosen discretisation. Consequently, the best choice for further
investigations of the theory on the lattice is the discretisation with the
smallest separation of the particles in the multiplet.

At $\beta=1.75$ we have done a reasonable amount of simulations with three,
instead of one, level of stout smearing. We have found that the fluctuations
of the lowest eigenvalues of the Hermitian Wilson-Dirac operator are
considerably reduced with the additional levels of stout smearing, but the
results for the mass spectrum are consistent with the data from one level of
smearing. Hence this modification represents rather a technical improvement
than a reduction of the lattice artefacts.

Symanzik's improvement program~\cite{Symanzik:1983dc,Symanzik:1983gh}
provides a systematic way to cancel the leading $O(a)$ lattice artefacts of
the Curci-Veneziano action. The fermion action is modified by the addition
of an irrelevant operator, the so-called clover term,
\begin{equation}
-  c_{sw} \frac{a}{4}\, \bar{\lambda}(x) \sigma_{\mu\nu} F^{\mu\nu} \lambda(x).
\label{clover}
\end{equation}

The coefficient $c_{sw}$ is tuned such that $O(a)$ lattice artefacts are
removed from on-shell quantities, like for instance masses of physical
particles~\cite{Luscher:1984xn,Sheikholeslami:1985ij,Wohlert:1987rf}. The
perturbative calculation of $c_{sw}$ has been presented up to $O(g^2)$ for
$\mathcal{N}=1$ SYM in Ref.~\cite{Musberg:2013foa}. It is, however, well
known that higher order corrections to $c_{sw}$ are non-negligible in the
range of gauge couplings used in practical Monte Carlo
simulations~\cite{Musberg:2013foa,Karavirta:2010ym}.

An alternative to the perturbative result is given by a mean-field rescaling
of the link variables with the fourth root of the plaquette, $u_0 = \langle
P \rangle^{1/4}$, due to a suppression of tadpole 
diagrams~\cite{Lepage:1992xa}. The resulting clover coefficient reads
\begin{equation}
c_{sw} = \frac{1}{u_0^3}\,.
\end{equation}

\begin{figure}[ht]
\centering
\subfigure[Chiral extrapolation.]%
{\includegraphics[width=0.46\textwidth]{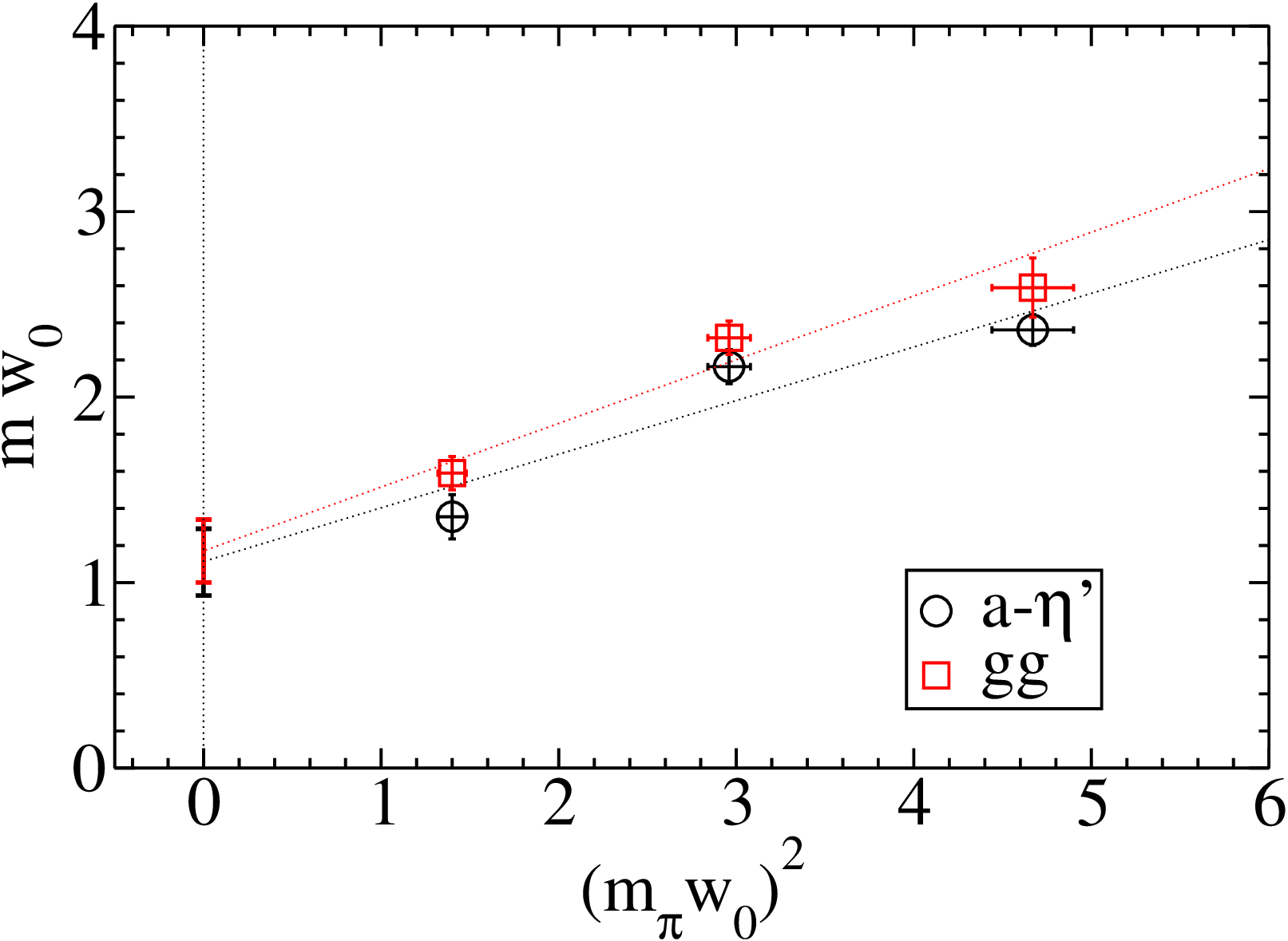}
\label{chiral_clover}}
\subfigure[Comparison between smeared (see Figure~\ref{f0etacontinuumcomp})
and clover improved fermions (yellow box).]%
{\includegraphics[width=0.48\textwidth]{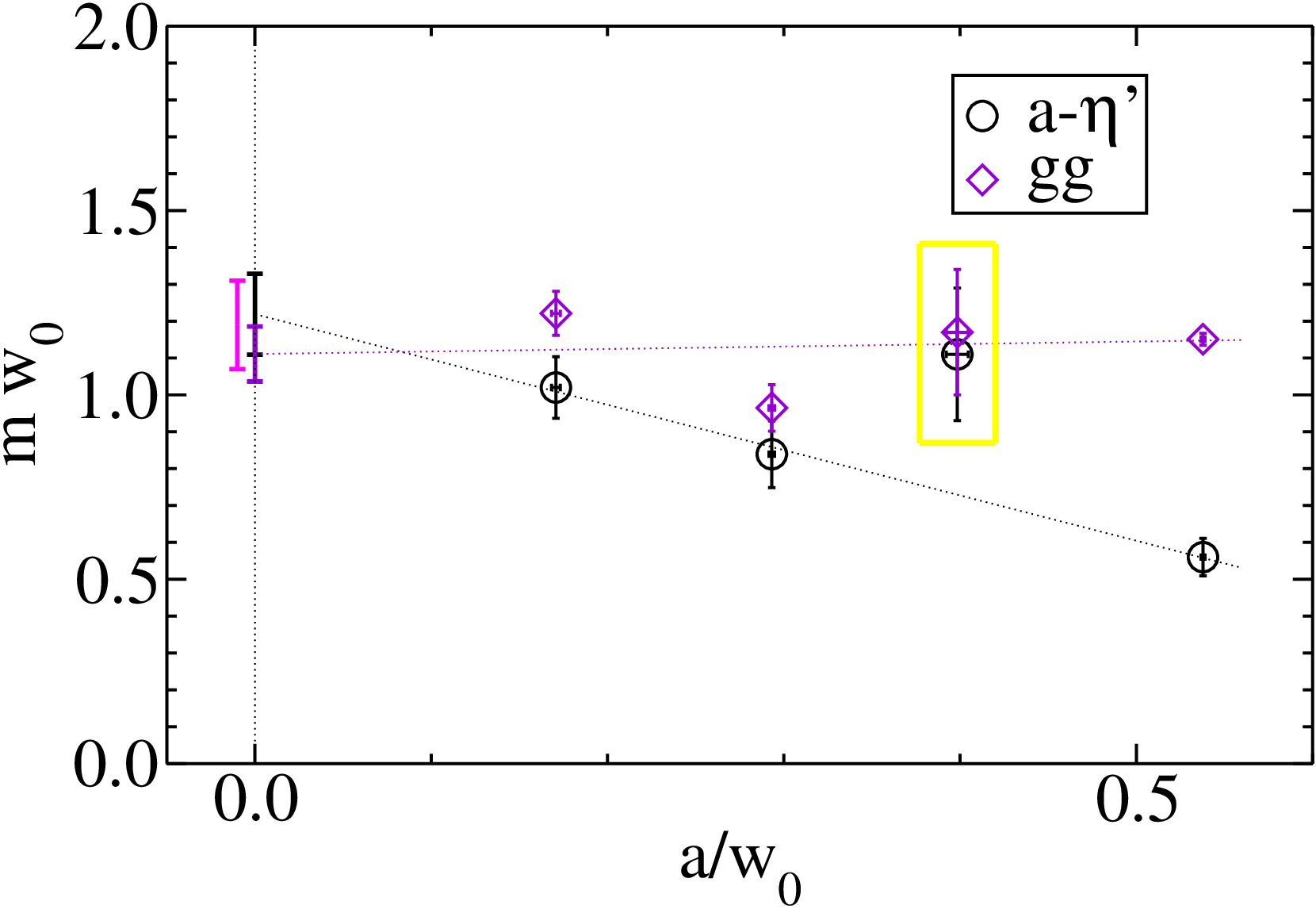}
\label{comparison_clover}}
\caption{a) Extrapolation of the $\aetap$ and gluino-glue masses to the
chiral limit computed with tadpole improved clover fermions. b) Comparison
of the extrapolations to the continuum limit of the $\aetap$ and gluino glue
masses with the tadpole clover improved action results (yellow box). Despite
the large lattice spacing, the results obtained with the improved action are
in agreement with the continuum limit extrapolation using the action with
one level of stout smearing on the fermions.}
\end{figure}

We have done some preliminary simulations of tadpole improved clover
fermions without stout smearing. The results are summarised in
Tab.~\ref{tab:chirallimitclover}. The chiral extrapolated value of $w_0$ is
determined to be $w_0=2.51(4)$. We measured the plaquette expectation value
at $\beta=1.7$ extrapolated to the chiral limit and, using the formula
above, we set $c_{sw}=1.467$. The limited statistics allows only for a
determination of the $\aetap$ and gluino-glue mass. The chiral extrapolation
is presented in Fig.~\ref{chiral_clover}. It is interesting to observe that
the values of the masses extrapolated to the chiral limit are already
compatible with their values extrapolated to the continuum limit using the
one-level stout smeared action, see Fig.~\ref{comparison_clover}.

%%%%%%%%%%%%%%%%%%%%%%%%%%%%%%%%%%%%%%%%%%%%%%%%%%%%%%%%%%%%%%%%%%%%%%%%
\section{Conclusions}

In this paper we have presented our latest results on the spectrum of bound
states in SYM. Reliable lattice simulations are now possible thanks to the
evolution of supercomputers and algorithms in recent years. Another
important step has been the localisation of the reliable parameters range
for the simulations which is limited by finite size effects and topological
freezing.

Based on these recent developments we have generated a large set of
configurations at different lattice spacings and gluino masses over several
years. At each lattice spacing we have performed an extrapolation of the
lowest bound state masses in the scalar, pseudoscalar, and spin-1/2 channel
to the chiral limit at zero gluino mass. As we have shown, this limit is
compatible with the restoration of the supersymmetric Ward identities. Our
last set of data at $\beta=1.9$ allows to complete the extrapolation to the
continuum limit.

The extrapolations to the continuum limit show agreement for the masses
within less than two standard deviations. This is consistent with the
formation of a SUSY multiplet, which is expected to contain a scalar and a
pseudoscalar boson in addition to the fermion.

There are two important conclusions that can be drawn from these
observations: a proper non-perturbative definition of the strongly
interacting supersymmetric theory is possible and there is no breaking of
SUSY in the low energy effective theory. This is equivalent to the absence
of an anomalous or spontaneous SUSY breaking in this theory.

Taking the unbroken SUSY in the continuum theory for granted, the second
conclusion is that the lattice method can be applied in a non-trivial
four-dimensional supersymmetric theory. SUSY, which is unavoidably broken by
the discretisation, is like Lorentz symmetry and chiral symmetry restored in
the continuum limit in SYM. In our case, this is achieved by a fine-tuning
of the hopping parameter.

In a first, short study we have shown that the breaking of SUSY indicated by
the mass splitting of the multiplet on a coarse lattice is significantly
reduced when a tadpole improved clover fermion action is used. In addition,
technical aspects of the methods are improved by stout smearing in the Dirac
operator. This indicates that clover improved fermions with stout smearing
seem to be the best choice for future lattice investigations of the theory.

In our present investigation we considered only the mass of the lightest
supermultiplet. A second chiral supermultiplet with higher mass has been
predicted~\cite{Farrar:1997fn,Farrar:1998rm,Feo:2004mr}. The determination
of the mass of the second supermultiplet on the lattice requires the
computation of the mixing between gluonic and fermionic operators. We plan
to investigate this aspect in the near future. The analysis of the mass of
the pseudoscalar glueball and of the excited state of the scalar glueball
indicates that mixing might be rather weak in the $0^{-+}$ channel. In this
case, the second excited supermultiplet appears to be roughly twice as
massive as the ground state.

%%%%%%%%%%%%%%%%%%%%%%%%%%%%%%%%%%%%%%%%%%%%%%%%%%%%%%%%%%%%%%%%%%%%%%%%
\section*{Acknowledgements}

The authors gratefully acknowledge the Gauss Centre for Supercomputing (GCS)
for providing computing time for a GCS Large-Scale Project on the GCS share
of the supercomputer JUQUEEN at J\"ulich Supercomputing Centre (JSC). GCS is
the alliance of the three national supercomputing centres HLRS
(Universit\"at Stuttgart), JSC (Forschungszentrum J\"ulich), and LRZ
(Bayerische Akademie der Wissenschaften), funded by the German Federal
Ministry of Education and Research (BMBF) and the German State Ministries
for Research of Baden-W\"urttemberg (MWK), Bayern (StMWFK) and
Nordrhein-Westfalen (MIWF). Further computing time has been provided on the
supercomputers JURECA and JUROPA at JSC, SuperMUC at LRZ, and on the compute
cluster PALMA of the University of M\"unster.

%%%%%%%%%%%%%%%%%%%%%%%%%%%%%%%%%%%%%%%%%%%%%%%%%%%%%%%%%%%%%%%%%%%%%%%%

%%%%%%%%%%%%%%%%%%%%%%%%%%%%%%%%%%%%%%%%%%%%%%%%%%%%%%%%%%%%%%%%%%%%%%%%
\begin{appendix}
\section{Tables}

\begin{table}[htbp]
  \begin{center}
    \begin{tabular}{|l|l|l|l|l|l|l|l|}
      \hline
      $\kappa$ & \# confs & $a m_{\api}$ & $a m_{\aetap}$ & $a m_{\afn}$ & $a m_{\glg}$ & $m_{gb} (0^{++})$& $m_{gb} (0^{-+})$ \\
      \hline
      0.1433  & 10374 & 0.28737(84) & 0.314(18) & 0.313(84) & 0.324(11)  & 0.35(2) & 0.60(3) \\
      \hline
      0.14387 & 10237 & 0.21410(33) & 0.260(11) & 0.265(53) & 0.295(10)  & 0.27(3) & 0.52(6) \\
      \hline
      0.14415 & 21090 & 0.17520(22) & 0.232(11) & 0.262(25) & 0.236(15)  & 0.29(2) & 0.52(4) \\
      \hline
      0.14435 & 10680 & 0.14129(59) & 0.203(12) & 0.227(19) & 0.2360(74) & 0.29(3) & 0.52(3) \\
      \hline
    \end{tabular}
  \end{center}
  \caption{Summary of the masses in lattice units for the one-level stout
           smeared action at $\beta=1.9$, lattice size $32^3 \times 64$.}
  \label{tab:chirallimit190}
\end{table}

\begin{table}[htbp]
  \begin{center}
    \begin{tabular}{|l|l|l|l|l|l|}
      \hline
       $\kappa$ & \# configs & $w_0/a$ & $a m_{\api}$ & $a m_{\aetap}$ & $a m_{\glg}$ \\
      \hline
      0.1600 & 4043 & 1.7456(87) & 0.8606(71) & 0.941(18)  & 1.030(46) \\
      \hline
      0.1620 & 3474 & 1.915(11)  & 0.6851(28) & 0.862(23) & 0.924(20)  \\
      \hline
      0.1640 & 1466 & 2.165(15)  & 0.4716(54) & 0.540(39)  & 0.633(26)  \\
      \hline
    \end{tabular}
  \end{center}
  \caption{Summary of the masses in lattice units for the tadpole clover
           improved action at $\beta=1.7$, $c_{sw}=1.467$, lattice size $16^3 
           \times 32$.}
  \label{tab:chirallimitclover}
\end{table}

\begin{table}[htbp]
%  \begin{center}
\hspace{-1cm}
    \begin{tabular}{|l|l|l|l|l|l|l|l|}
      \hline
       $\beta$ & $a m_{\aetap}$ & $a m_{\afn}$  & $a m_{\glg}$ & $a m_{gb} (0^{++})$ &  
       $a m_{gb} (0^{-+})$ & $w_0/a$ & $r_0/a$ \\
      \hline
      1.90 & 0.174(14) & 0.201(35) & 0.208(10) & 0.253(26)  & 0.486(35) & 5.858(84) & 13.95(12) \\
      \hline
      $1.75^{l=1}$ & 0.246(26) & 0.228(71) & 0.283(18) & 0.294(62) & 0.63(22) & 3.411(18) & 9.47(14) \\
      \hline
      $1.75^{l=3}$ & 0.268(22)  & 0.331(30)  & 0.3295(83)  & 0.393(33)  & 1.014(20) &  3.189(11)  & 9.28(7)   \\
      \hline
      1.60 & 0.301(27) & 0.256(81) & 0.619(86) & 0.51(13)  & -- & 1.8595(39) & 5.78(15) \\
      \hline
    \end{tabular}
%  \end{center}
  \caption{Summary of the masses and scale parameters in lattice units
           extrapolated to the chiral limit ($\kappa=\kappa_c$). For
           $\beta=1.75$ we report the values for one ($l=1$) and three
           ($l=3$) levels of stout smearing.}
  \label{tab:chirallimit}
\end{table}

\begin{table}[htbp]
  \begin{center}
    \begin{tabular}{|l|l|l|l|l|}
      \hline
          &  $s m_{\aetap}$ & $s m_{\afn}$  & $s m_{\glg}$ & $s m_{gb} (0^{++})$ \\
       \hline
       $s=r_0$ &  2.97(33) &  3.67(85) & 2.23(23) & 3.96(83) \\
      \hline
       $s=w_0$ &  1.22(11) & 1.43(28) & 1.111(74) & 1.67(25)   \\
      \hline
       $s=w_0^*=r_0/(r_0/w_0)$  &  1.34(17) & 1.66(40) & 1.01(12) & 1.79(39)   \\
      \hline
       $(w_0^*-w_0)/\Delta{w_0}$  &  1.09 & 0.82 & 1.36 & 0.48 \\
      \hline
    \end{tabular}
  \end{center}
  \caption{Summary of the masses extrapolated to the continuum limit using
           different scales. $w_0^*$ is a parameter determined from $r_0$
           using the ratio $r_0/w_0=2.21(12)$. In the last row the
           comparison between $w_0$ and $w_0^*$ is shown. In this table
           the mass of the glueball $0^{-+}$ is not present because we
           have not enough data for the extrapolatation to the continuum
           limit.}
  \label{tab:continuumlimit}
\end{table}

\end{appendix}

\end{document}